\documentclass[aps,pra,twocolumn,superscriptaddress,showpacs]{revtex4}
\usepackage{dcolumn}                 
\newcolumntype{w}[1]{D{.}{.}{#1}}
\newcommand*{\centt}[1]{\multicolumn{1}{c}{#1}}
\newcommand*{\cent}[1]{\multicolumn{1}{c}{$#1$}}
\usepackage{times}
\usepackage{nicefrac}
\usepackage{amsmath}
\usepackage{amsfonts}
\usepackage{amssymb}
\usepackage{amsthm}

\usepackage{color}

\begin{document}
\preprint{Version 1.1}

\title{Ground--state hyperfine splitting in the Be$^+$ ion}

\author{Mariusz Puchalski}
\affiliation{Faculty of Physics, University of Warsaw, Ho{\.z}a 69, 00-681 Warsaw, Poland}
\affiliation{Faculty of Chemistry, Adam Mickiewicz University,
             Grunwaldzka 6, 60-780 Pozna{\'n}, Poland}

\author{Krzysztof Pachucki}
\affiliation{Faculty of Physics, University of Warsaw, Ho{\.z}a 69, 00-681 Warsaw, Poland}

\begin{abstract}
Relativistic and QED corrections are calculated for the hyperfine splitting
(hfs) in the $2S_{1/2}$ 
ground state of $^{9}$Be$^+$ ions with an exact account for electronic correlations. 
The achieved accuracy is sufficient to determine the finite nuclear size effects
from the comparison to the experimental hfs value.
The obtained results establish the ground to determine the neutron halo in  $^{11}$Be.
\end{abstract}

\maketitle
\section{Introduction}
High-precision atomic spectroscopy makes possible the accurate determination of the electromagnetic 
properties of nuclei, including short-lived exotic isotopes. The best known example is the mean square
nuclear charge radius, which can be obtained from the isotope shift of atomic
transition energies \cite{litiso}. 
Here, we develop a computational technique for the determination 
of magnetic properties of nuclei, which can be obtained from hyperfine splitting. 
Apart from the magnetic moment, they are very much unknown. 
The atomic hyperfine splitting is sensitive to the distribution of the magnetic moment and, to some extent,
to combined polarizabilities. Altogether it can be expressed in terms of the effective Zemach radius $\tilde{r}_Z$ \cite{lit_hfs}.
The results recently obtained for $^{6,7}$Li indicate that $\tilde{r}_Z(^6$Li$)$ is more than 40\% smaller than $\tilde{r}_Z(^7$Li$)$,
which is not necessarily easy to understand. This significant difference can probably
be resolved only by detailed nuclear structure calculations.  

In this work we perform analogous, accurate calculations of hyperfine splitting in the $2S_{1/2}$ 
ground state Be$^+$ ions, in order to determine  $\tilde{r}_Z$ for
$^{7,9,11}$Be isotopes. Since the magnetic moment is known for $^9$Be, we
can compare $\tilde{r}_Z(^9{\rm Be})$ with theoretical predictions here. 
For other $^{7,11}$Be isotopes our calculations lay the foundation for the determination
of $\tilde{r}_Z$, once the magnetic moment is experimentally known.
It would be very interesting to confirm the large neutron halo in $^{11}$Be
using atomic spectroscopy measurements, and to study
the dependence of Bohr-Weisskopf effects on the isotope.

Our computational approach is based on
explicitly correlated basis functions. This allows us to accurately solve
the Sch\"odinger equation, while relativistic and QED effects are calculated perturbatively 
in terms of expectation values with the nonrelativistic wave function.

\section{Effective Hamiltonian}
Hyperfine splitting is a result of the interaction between the nuclear magnetic moments of atomic nuclei and 
electrons.
In the nonrelativistic QED approach, relativistic and QED corrections are expressed in terms
of an effective Hamiltonian, so
the expansion in the fine structure constant $\alpha$ is of the form
\begin{eqnarray}
E_{\rm hfs} &=& \langle H^{(4)}_{\rm hfs}\rangle +\langle H^{(5)}_{\rm hfs}\rangle  \label{Ehfs} \\  
&&  +2\,\langle H^{(4)}\,\frac{1}{(E-H)'}\,H^{(4)}_{\rm hfs}\rangle + \langle H^{(6)}_{\rm hfs}\rangle
 + \langle H^{(6)}_{\rm rad}\rangle \nonumber \\ 
 && + \langle H^{(7)}_{\rm hfs}\rangle +\cdots\,. \nonumber 
\end{eqnarray}
where $ H^{(n)} \sim m \alpha^n$, and the nonrelativistic Hamiltonian in the clamped nucleus limit
and the nonrelativistic energy of the ground state are $H$ and $E$, respectively. 
Higher order terms, denoted by dots, are neglected as the highest order
term $H^{(7)}_{\rm hfs}$ will be calculated in an approximate way.

\subsection{Leading order hfs}
The leading interaction $H^{(4)}_{\rm hfs}$  of order $m\,\alpha^4$ 
between the nuclear spin $\vec I$ and electron spins $\vec \sigma_a$
is obtained from the nonrelativistic coupling of electrons to the electromagnetic field
\begin{equation}
H^{(4)}_{\rm hfs} = -\sum_a \frac{e}{m}\,\vec p_a\cdot\vec A(\vec r_a) 
-\frac{e}{2\,m}\,\frac{g}{2}\,\vec\sigma_a\cdot\vec B(\vec r_a)
\label{02}
\end{equation}
with $\vec A$ and $\vec B$ fields derived from the nuclear magnetic
moment $\vec \mu$
\begin{eqnarray}
e\,\vec A(\vec r) &=& \frac{e}{4\,\pi}\,\vec\mu\times\frac{\vec r}{r^3} = 
-Z\,\alpha\,\frac{g_N}{2\,M}\,\vec I\times\frac{\vec r}{r^3}\\
e\,B^i(\vec r) &=& \bigl(\nabla\times\vec A\bigr)^i 
= -Z\,\alpha\,\frac{g_N}{2\,M}\,\frac{8\,\pi}{3}\,\delta^3(r)\,I^i \nonumber \\
&& +Z\,\alpha\,\frac{g_N}{2\,M}\,\frac{1}{r^3}\,\biggl(\delta^{ij}-
3\,\frac{r^i\,r^j}{r^2}\biggr)\,I^j
\end{eqnarray}
After some simplifications, Eq. (\ref{02}) becomes 
\begin{eqnarray}
H^{(4)}_{\rm hfs} &=& \varepsilon\,
\Bigl(\frac{g}{2}\,H^{A}_{\rm hfs}+H^{B}_{\rm hfs}+H^{C}_{\rm hfs}\Bigr)\,,\label{H4hfsABC}\\
H^{A}_{\rm hfs} &\equiv& \sum_a \vec{I}\cdot\vec{\sigma}_a H^{A}_{a,\rm hfs} = \sum_a\,\frac{4\,Z\,\alpha}{3\,m^3}\,
\vec{I}\cdot\vec{\sigma}_a\,\pi\,\delta^3(r_a)\,,\label{06} \\
H^{B}_{\rm hfs} &\equiv& \vec{I}\cdot \vec{H}^{B}_{\rm hfs} = \sum_a\,\frac{Z\,\alpha}{m^3}\,
\vec{I}\cdot\frac{\vec{r}_a\times\vec{p}_a}{r_a^3}\,,\\
H^{C}_{\rm hfs} &\equiv& I^i\,\sigma_a^j\,H^{Cij}_{a,\rm hfs} \\
&=& \sum_a\,-\frac{Z\,\alpha}{2\,m^3}\,\frac{I^i\,\sigma_a^j}{r_a^3}\,
\biggl(\delta^{ij}-3\,\frac{r_a^i\,r_a^j}{r_a^2}\biggr)\,,\nonumber 
\end{eqnarray}
where
\begin{equation}
\varepsilon = \frac{m^2}{M}\,\frac{g_{\rm N}}{2}\,,
\end{equation}
$M$ ($m$) are masses and $g_{\rm N}$ ($g$) are $g$-factors of the nucleus (electron). 
The relation of $g_{\rm N}$ to the magnetic moment $\mu$ of the nucleus with charge $Z$ is 
\begin{equation}
g_{\rm N} = \frac{M}{Z\,m_{\rm p}}\,\frac{\mu}{\mu_{\rm N}}\,\frac{1}{I}
\end{equation}
where $\mu_{\rm N}$ is the nuclear magneton and $I$ is the nuclear spin. 
The only nonvanishing term in the ground state $H^{A}_{\rm hfs}$ is well known as the Fermi contact interaction.
Expectation values of $H^{B}_{\rm hfs}$ and $H^{C}_{\rm hfs}$ contribute in the second-order of 
perturbation calculus. In principle these terms also involve the electron $g-$factor but here we set $g=2$.

\subsection{Correction of order \boldmath{$m\,\alpha^5$}}

Correction $\langle H^{(5)}_{\rm hfs} \rangle$ of order $m\,\alpha^5$
is a Dirac-$\delta$ interaction with the coefficient
obtained from the two-photon forward scattering amplitude.
It has the same form as in hydrogen and depends
on the nuclear structure. At the limit of a point spin $1/2$ nucleus it is
\begin{equation}
H^{(5)}_{\rm hfs}  =  -H^A_{\rm hfs}\,\frac{3\,Z\,\alpha}{\pi}\,
\frac{m}{m_{\rm N}}\,\ln\Bigl(\frac{m_{\rm N}}{m}\Bigr) \equiv H^{(5)}_{\rm rec}\label{11}
\end{equation}
a small nuclear recoil correction. For a finite-size nucleus $H^{(5)}_{\rm hfs}$ does not vanish
at the non-recoil limit. 
When assuming a heavy and rigid nucleus, $H^{(5)}_{\rm hfs}$ takes the form
\begin{equation}
H^{(5)}_{\rm hfs} =  \varepsilon\,H^A_{\rm hfs}\,(-2\,Z\,\alpha\,m\,r_Z) \label{12}
\end{equation}
where
\begin{equation}
r_Z = \int d^3r\,d^3r'\,\rho_E(r)\,\rho_M(r')\,|\vec r-\vec r'|
\end{equation}
and $\rho_E$ and $\rho_M$ are electric charge and magnetic moment density.
The inelastic contribution, usually neglected,
is sometimes important. Since it depends on 
nuclear excitations, this correction is very difficult to estimate
and usually limits the precision of theoretical predictions.
For this reason, we will interpret our calculation using experimental values
as a determination of the effective Zemach radius according to
Eq. (\ref{12}), where $r_Z$ is replaced by  $\tilde{r}_Z$

\subsection{Relativistic correction of order \boldmath{$m \alpha^6$}}

The first term for the  relativistic correction of order $m \alpha^6$ in Eq.~\eqref{Ehfs} 
comes from a perturbation of the wave function by the Breit-Pauli Hamiltonian $H^{(4)}$  in the non-recoil limit
\begin{eqnarray}
H^{(4)} &=& H^{A}+H^{B}+H^{C}\,,\label{H4ABC}\\
H^{A} &\equiv& \sum_a\, H^{A}_a = \sum_a\,\left[ -\frac{p_a^4}{8\,m^3}+
\frac{Z\,\alpha\,\pi}{2\,m^2}\,\delta^3(r_a)\right] \\ 
&& \hspace{-0.5cm} +\sum_{a > b}\left[
\frac{\pi\,\alpha}{m^2}\,\delta^3(r_{ab})
-\frac{\alpha}{2\,m^2}\,p_a^i\biggl(
\frac{\delta^{ij}}{r_{ab}}+\frac{r^i_{ab}\,r^j_{ab}}{r^3_{ab}}\biggr)\,p_b^j\right], \nonumber \\
H^{B} &\equiv& \sum_a\, \vec \sigma_a \cdot \vec H^{B}_a = \sum_a\,\frac{Z\,\alpha}{4\,m^2}\,
\frac{\vec{r}_a}{r_a^3}\times\vec{p}_a\cdot\vec{\sigma}_a \\
&& +\sum_{a\neq b}\, \frac{\alpha}{4\,m^2}\,
\frac{\vec{r}_{ab}}{r_{ab}^3}\times(2\,\vec{p}_b-\vec p_a)\cdot\vec{\sigma}_a \,, \nonumber \\
H^{C} &\equiv& \sum_{a>b} \sigma_a^i\,\sigma_b^j \,H^{Cij}_{ab} \\
&=& \sum_{a>b} \frac{\alpha}{4\,m^2}\,\frac{\sigma_a^i\,\sigma_b^j}{r_{ab}^3}\biggl(
\delta^{ij}-\frac{3\,r_{ab}^i\,r_{ab}^j}{r_{ab}^2}\biggr)\,. \nonumber 
\end{eqnarray}
The next term, $H_{\rm hfs}^{(6)}$, includes nuclear spin-dependent operators
that contribute at order $m\,\alpha^6$. In hydrogenic systems it leads to the so-called
Breit correction. For three-electron atoms this term 
was first derived in Ref. \cite{krp_hfs} and recently rederived and simplified
in \cite{lit_hfs}. The result is
\begin{eqnarray}
H^{(6)}_{\rm hfs} &=& \varepsilon\,
\sum_a\, \vec\sigma_a\cdot\vec I\,
\biggl[\frac{(Z\,\alpha)^2}{6\,m^4}\,\frac{1}{r_a^4}
-\frac{Z\,\alpha}{12\,m^5}\,
\bigl\{p_a^2\,,\,4\,\pi\,\delta^3(r_a)\bigr\} \nonumber \\
&& + \sum_{b\neq a}
\frac{Z\,\alpha^2}{6\,m^4}\,\frac{\vec r_{ab}}{r_{ab}^3}\cdot
\biggl(2\,\frac{\vec r_b}{r_b^3}-\frac{\vec r_a}{r_a^3}\biggr)\biggr],
\label{H6hfs}
\end{eqnarray}
where braces denote an anticommutator.
The resulting operators are divergent, and in the next section we describe 
the cancellation of singularities with those in second-order matrix elements.

\subsection{Radiative corrections of order \boldmath{$m \alpha^{6,7}$}}

$H^{(6)}_{\rm rad}$ in Eq. (\ref{Ehfs}) is a QED radiative correction \cite{nist, eides}
\begin{equation}
H^{(6)}_{\rm rad} =  H^A_{\rm hfs}\,\alpha\,(Z\,\alpha)\,\biggl(\ln 2-\frac{5}{2}\biggr)\,,
\label{23}
\end{equation}
which is similar to that in hydrogen. There are no further corrections
of this order, so all terms at $m\,\alpha^6$ are known exactly.

The last term $E^{(7)}_{\rm hfs}$ of order $m\,\alpha^7$ is calculated approximately using the hydrogenic
value for the one-loop correction from \cite{UJ} and the two-loop
correction from \cite{eides},
\begin{eqnarray}
H^{(7)}_{\rm hfs} &=& H^A_{\rm hfs}\,\biggl[
\frac{\alpha}{\pi}\,(Z\,\alpha)^2\,
\biggl(-\frac{8}{3}\,\ln^2(Z\,\alpha)
\nonumber \\ &&
+ a_{21}\,\ln(Z\,\alpha) + a_{20}\biggr)
+ \frac{\alpha^2}{\pi}\,(Z\,\alpha)\,b_{10}\biggr]\,, \label{24}
\end{eqnarray}
where $a_{21}(2S) = -1.1675$, $a_{20}(2S) = 11.3522$ and $b_{10} = 0.771\,652$.

\subsection{Hyperfine structure constant}

The hyperfine splitting can be expressed in terms of the hyperfine constant $A$
\begin{equation}
E_{\rm hfs} = \vec I\cdot\vec J\,A
\end{equation}
where $\vec J$ is the total electronic angular momentum. If we use the notation that 
$H_{\rm hfs} = \vec I\cdot\vec H_{\rm hfs}$, then
\begin{equation}
A = \frac{1}{J\,(J+1)}\,\langle \vec J\cdot\vec H_{\rm hfs}\rangle\,.
\end{equation}
The expansion of $A$ in $\alpha$ takes the form
\begin{equation}
A = \varepsilon\,\bigg(\frac{g}{2}\,\alpha^4 A^{(4)} + \sum_{n=5}^\infty\alpha^n\,A^{(n)}\bigg)
\end{equation}
All the results of numerical calculations will be presented here in terms of dimensionless coefficients~$A^{(n)}$. 
The leading order term $A^{(4)}$ obtained form Eq.~\eqref{06} is in turn
expanded in the  reduced electron mass $\mu$  to the nuclear mass $M$ ratio
\begin{eqnarray}
A^{(4)} &=&\frac{1}{J\,(J+1)}\frac{4\,\pi\,Z}{3}\, \bigg \langle \vec J\cdot\vec\sigma_a\, \bigg[ \delta^3(r_a) \label{A4} \\
&& -\frac{\mu}{M} \Big( 3\,\delta^3(r_a) + 2\,[\delta^3(r_a)]_{\rm mp} \Big)
  \bigg] \bigg \rangle\,. \nonumber\\
       &=& A^{(4,0)} - \frac{\mu}{M}\,A^{(4,1)}\,.
\end{eqnarray}
The finite mass correction due to mass scaling of the $\delta^3(r_a)$ operator is included into $A^{(4,1)}$, 
as well as the second-order element with the mass polarization correction to the wave function
\begin{equation}
[\delta^3(r_a)]_{\rm mp} = \delta^3(r_a) \frac{1}{(H-E)'} \, \sum_{b>c} \vec p_b \cdot \vec p_c.
\end{equation}
The next to leading correction $A^{(5)}$ and all others
are obtained in the leading order in the mass ratio, so that
\begin{eqnarray}
A^{(5)}_{\rm rec} &=& -A^{(4)}\,
\frac{3\,Z}{\pi}\,\frac{m}{m_{\rm N}}\,\ln\Bigl(\frac{m_{\rm N}}{m}\Bigr)\,.\label{32}\\
A^{(5)} &=& A^{(4)}\,(-2\,Z\,m\,\tilde{r}_Z)
\end{eqnarray}
The most demanding part of the calculation is the correction of order $m \alpha^6$
given by $A^{(6)}$. Due to the symmetry of intermediate states in 
the second-order matrix element of Eq.~\eqref{Ehfs}, the $A$, $B$ and $C$ parts of the hfs Hamiltonian 
give the non-vanishing contributions with the corresponding $A$, $B$ and
$C$ parts of Eq.~\eqref{H4ABC}.  
Of note, the matrix element of the first-order term in Eq.~\eqref{H6hfs} and 
the second-order $A$ terms are divergent separately at small $r_a$. However, these divergences can be 
eliminated in the sum of both terms denoted by
$A^{(6)}_{AN}$. So the complete $A^{(6)}$ is of the form
\begin{eqnarray}
A^{(6)} &=& A^{(6)}_{AN} + A^{(6)}_B + A^{(6)}_C + A^{(6)}_R \\
A^{(6)}_{AN} &=& \frac{2}{J\,(J+1)}\,\biggl\langle \sum_a
\vec J \cdot \vec \sigma_a\,H^A_{a,\rm hfs}\,\frac{1}{(E-H)'}\,H^A\biggr\rangle \nonumber \\&&
+\frac{1}{J\,(J+1)}\,\bigg\langle\vec J\cdot\vec\sigma_a\,\bigg[
\frac{Z^2}{6}\,\frac{1}{r_a^4} - \frac{2\,Z}{3}\,p_a^2\,\pi\,\delta^3(r_a) \nonumber \\
&&
+ \sum_{b\neq a}\frac{Z}{6}\,\frac{\vec r_{ab}}{r^3_{ab}}\cdot\Bigl(2\,\frac{\vec r_b}{r^3_b} 
- \frac{\vec r_a}{r^3_a}\Bigr)\bigg]\bigg\rangle \label{A6AN}\\
A^{(6)}_{B} &=& \frac{2}{J\,(J+1)}\,\biggl \langle \vec J\cdot \vec H^B_{\rm hfs}\,
\frac{1}{(E-H)'}\,H^B\biggr\rangle \label{A6B}\\
A^{(6)}_{C} &=&\frac{2}{J\,(J+1)}\,\biggl\langle \sum_a J^i\,\sigma_a^j\,H^{Cij}_a
\,\frac{1}{(E-H)'}\,H^C\biggr\rangle  \nonumber\\ \label{A6C} \\
A^{(6)}_{R} &=&A^{(4)}\,\Bigl(\ln 2-\frac{5}{2}\Bigr). 
\end{eqnarray}
And the higher order term is
\begin{eqnarray}
A^{(7)} &=& A^{(4)}\,\biggl[ \frac{Z^2}{\pi}\,\biggl(-\frac{8}{3}\,\ln^2(Z\,\alpha) + a_{21}\,\ln(Z\,\alpha) + a_{20}\biggr) \nonumber \\ 
 &&+ \frac{Z}{\pi}\,b_{10}\biggr]
\end{eqnarray}

\section{Calculations}
\subsection{Cancellation of singularities in \boldmath{$A^{(6)}_{AN}$}}

The operators in Eq.~\eqref{A6AN} are transformed with the use of
\begin{eqnarray}
4\,\pi\,\delta^{3}(r_a)&\equiv& 4\,\pi\,[\delta^{3}(r_a)]_r - \sum_a\,\biggl\{\frac{2}{r_a}\,,\,E-H\biggr\}\,, \label{drared1} \\ 
H^A &\equiv& [H^A]_r +
\frac{1}{4}\,\sum_a\,\biggl\{\frac{Z}{r_a}\,,\,E-H\biggr\} \,. \label{HAred1}
\end{eqnarray}
Regularized operators $[H^{A}]_r$ and $[\delta^{3}(r_a)]_r$ have exactly the same expectation value 
as the operator inside the square brackets if the exact wave function is used. From Eq.~\eqref{drared1} and Eq. \eqref{HAred1} 
we can obtain the following formulas
\begin{eqnarray}
4 \pi\,[\delta (r_a)]_r &=& 4\,(E - V) \,\frac{1}{r_a} - 2 \sum_{b} \vec p_b \,\frac{1}{r_a}\, \vec p_b \,, \label{drared2}\\
{}[H^{A}]_r &=&  \sum_a \bigg[-\frac{1}{8}\, [p_a^4]_r 
+ \frac{1}{2}\,p_a^i \bigg( V + \frac{Z}{2} \sum_b\frac{1}{r_b}  \bigg)\,p_a^i \bigg] \nonumber\\
&& \hspace{-0.5cm} -\bigg( V + \frac{Z}{2} \sum_b\frac{1}{r_b}  \bigg) (E-V)  \\
&& \hspace{-0.5cm} +\sum_{a > b}\left[ 3 \pi\,\delta^3(r_{ab}) -\frac{1}{2}\,p_a^i\biggl(
\frac{\delta^{ij}}{r_{ab}}+\frac{r^i_{ab}\,r^j_{ab}}{r^3_{ab}}\biggr)\,p_b^j\right], \nonumber \\
\sum_a [p^4_a]_r &=& 4 \,(E - V)^2 - 2\, \sum_{a > b} \vec p_a^{\,2}\,\vec p_b^{\,2} \,,
\end{eqnarray}
After this transformation both the first and second-order matrix elements in $A^{(6)}_{AN}$ 
become separately finite
\begin{eqnarray}
A^{(6)}_{AN} &=& A^{(6)}_{A} + A^{(6)}_{N} \\
A^{(6)}_{A} &=& 
\frac{2}{J\,(J+1)}\,\biggl\langle \sum_a
\vec J \cdot \vec \sigma_a\,[H^A_{a,\rm hfs}]_r \,\frac{1}{(E-H)'}\,[H^A]_r \biggr\rangle  \nonumber \\
A^{(6)}_{N}&=&\frac{1}{J\,(J+1)}\,\bigg\langle\vec J\cdot\vec\sigma_a\,\frac{Z}{6}\,
\bigg[\frac{1}{r_a}\,\sum_b\,p_b^4 - 4\,\pi\,\delta^3(r_a)\,p_a^2 \nonumber  \\
&& \hspace{-1cm} + \sum_{b\neq a}\frac{\vec r_{ab}}{r_{ab}^3} \cdot \biggl(2\frac{\vec r_b}{r_b^3}-\frac{\vec r_a}{r_a^3}\biggr)  
+4\,\pi\,Z\,\sum_{b\neq a} \biggl(\frac{\delta^3(r_a)}{r_b}-\frac{\delta^3(r_b)}{r_a}\biggr) \nonumber  \\
&& \hspace{-1cm} -\frac{2}{r_a}\,\sum_{b>c} 4\,\pi\,\delta^3(r_{bc})  +4\,\sum_{b>c}p_b^i\,\frac{1}{r_a}\,\biggl(
\frac{\delta^{ij}}{r_{bc}}+\frac{r_{bc}^i\,r_{bc}^j}{r_{bc}^3}\biggr)\,p_c^j \nonumber  \\
&& \hspace{-1cm} -4\,\pi\,Z\,\delta^3(r_a)\,\biggl\langle\sum_b\,\frac{1}{r_b}\biggr\rangle
+\frac{8}{r_a}\,\langle H^A\rangle\bigg] \bigg\rangle \label{A6AN-2}
\end{eqnarray}
It is still necessary, however, to remove apparent singularities in some of the first-order operators
by repeated use of the Schr\"odinger equation, namely 
\begin{eqnarray}
\biggl\langle\frac{1}{r_a}\,\sum_b\,p_b^4-4\,\pi\,\delta^3(r_a)\,p_a^2\biggr\rangle &=&
\biggl\langle -2\,\sum_{b; b\neq a}\,\frac{\vec r_{ab}}{r_{ab}^3}\cdot\frac{\vec r_a}{r_a^3} \\
&& \hspace*{-30ex} +\frac{4}{r_a}\,\biggl[(E-V)^2-\frac{Z^2}{r_a^2}\biggr]
    -2\,\sum_{b,c; b>c} p_b^2\,\frac{1}{r_a}\,p_c^2 
 +2\,Z\,\vec p_a\,\frac{1}{r_a^2}\,\vec p_a \nonumber \\
 && \hspace{-4.5cm} +\biggl[8\,\pi\,\delta^3(r_a)+\frac{4\,Z}{r_a^2}\biggr]\,\biggl(\sum_{b; b \neq a}
     \frac{p_b^2}{2}+V+\frac{Z}{r_a}-E\biggr)\biggr\rangle\nonumber 
\end{eqnarray}
In this form, all matrix elements  with the nonrelativistic wave function can safely be calculated.

\subsection{Wave function and first-order operators}

The wave function for a lithium-like system in the ground state is represented as a linear combination of $\psi$ terms,
the antisymmetrized product of $S$-symmetry spatial $\phi$ and doublet spin functions \cite{lit_wave}
\begin{equation}
\psi = \frac{1}{\sqrt{6}}\,{\cal A}\big[\phi(\vec r_1,\vec r_2,\vec r_3)\,
[\alpha(1)\,\beta(2)-\beta(1)\,\alpha(2)]\,\alpha(3)\big]\,,\label{wf_psiS} 
\end{equation}
where ${\cal A}$ denotes antisymmetrization with respect to electron variables, and the spin
functions are defined by  $\sigma_z\,\alpha(.) = \alpha(.) $ and $\sigma_z\,\beta(.) = -\beta(.) $.
\begin{table}[!hbt]
\renewcommand{\arraystretch}{1.0}
\caption{Symmetrization coefficients in matrix elements}
\label{table2}
\begin{ruledtabular}
\begin{tabular}{lrrrrr}
$(k,l,m)$  & $c_{klm}$ & $c^A_{klm}$ & $c^{F1}_{klm}$ & $c^{F2}_{klm}$ & $c^{F3}_{klm}$  \\
\hline 
 \\
 $(1,2,3)$    &   2   &   1 &    0    &    0    &     2   \\
 $(1,3,2)$    &  -1   &  -1 &    1    &   -1    &    -1   \\
 $(2,1,3)$    &   2   &   1 &    0    &    0    &     2   \\
 $(2,3,1)$    &  -1   &  -1 &   -1    &    1    &    -1   \\
 $(3,1,2)$    &  -1   &   1 &    1    &   -1    &    -1   \\
 $(3,2,1)$    &  -1   &  -1 &   -1    &    1    &    -1   \\
  \end{tabular}
\end{ruledtabular}
\end{table}

Until now, the most accurate nonrelativistic wave functions for 
lithium-like systems were obtained using the Hylleraas-type basis functions \cite{king_lit,yan_lit,Wang_2012,lit_wave,slater1} with elements 
of the form
\begin{equation}
\phi_H(\vec r_1, \vec r_2, \vec r_3) =
r_{23}^{n_1}\,r_{31}^{n_2}\,r_{12}^{n_3}\,r_{1}^{n_4}\,r_{2}^{n_5}\,r_{3}^{n_6}
\,e^{-\alpha_1\,r_1-\alpha_2\,r_2-\alpha_3\,r_3},
\end{equation}
with nonnegative integers $n_k$. We use the wave function obtained variationally 
\cite{slater1}, to evaluate most of the first-order matrix elements of the hyperfine structure operators, 
including the Fermi contact term in Eq.~\eqref{A4}. Such matrix elements can be expressed as a linear combination of Hylleraas integrals, defined as
\begin{eqnarray}
f(n_1,n_2,n_3,n_4,n_5,n_6) &=& \int \frac{d^3 r_1}{4\,\pi}\,
                               \int \frac{d^3 r_2}{4\,\pi}\,
                               \int \frac{d^3 r_3}{4\,\pi}\, \nonumber \\
&& \hspace{-1.5cm} \times \, r_{23}^{n_1-1}\,r_{31}^{n_2-1}\,r_{12}^{n_3-1}\,r_{1}^{n_4-1}\,r_{2}^{n_5-1}\,r_{3}^{n_6-1}\, \nonumber\\
&& \hspace{-1.5cm} \times \, e^{-w_1\,r_1-w_2\,r_2-w_3\,r_3} 
\end{eqnarray}
In a series of papers, we have formulated an analytic method for calculations of Hylleraas integrals with 
recursion relations \cite{recursions,recursions2,lit_wave}, which is sufficient for the
evaluation of energy levels including corrections up to $m \alpha^5$ order \cite{lit_rel,dlines}. 
At higher orders, additional classes of Hylleraas integrals are necessary, e.g. $f(-1,-1,n_3,n_4,n_5,n_6)$.
These difficult integrals have been solved with the use of Neumann-type expansions \cite{king_int},
but this approach is not effective enough in large-scale calculations. 
There is also an exceptional group of operators of $A^{(6)}_N$ with accompanying Dirac-$\delta$ operators. 
We are not able to regularize them by rewriting in a form analogous to
Eq.~\eqref{drared2}. However, the direct treatment of Dirac-$\delta$ is applicable
in the Hylleraas basis set, where the matrix elements are expressed in terms of
well-known two-electron integrals \cite{gamma1,gamma2,gamma3}.

It has been demonstrated recently \cite{lit_hfs} that matrix elements of
some complicated operators, which are intractable in the Hylleraas basis,
can be calculated with exponentially correlated Gaussian (ECG) functions \cite{mitroy2013}
\begin{equation}
\phi_G(\vec r_1, \vec r_2, \vec r_3) = \,e^{-\alpha_1\,r_1^2-\alpha_2\,r_2^2-\alpha_3\,r_3^2
-\beta_1\,r_{23}^2-\beta_2\,r_{13}^2-\beta_3\,r_{12}^2}\,.
\end{equation}
Even if the wave function in the ECG basis decays too fast at long inter-particle distances and 
fails to correctly satisfy the Kato cusp condition, it can be
sufficiently accurate to obtain matrix elements of these complicated hfs operators with at 
least 4-5-digit precision (see numerical results in Table~\ref{table3}). 
This has been verified by more accurate calculations 
using correlated Slater functions \cite{slater1,slater2} for the lithium case \cite{lit_hfs}. 
Hopefully, the numerically dominating operators in  $A_N^{(6)}$ are those obtained 
with Hylleraas functions. This is especially important due to the cancellation
of about 2-3 digits in this sum. 
Calculations of mean values with ECG basis for operators like $\frac{\vec r_{ab}\cdot\vec r_a}{r_{ab}^{3}r_a^{3}}$, 
$p_b^2\frac{1}{r_a}p_c^2$ or $p_b^i\frac{1}{r_a}(\frac{\delta^{ij}}{r_{bc}}+\frac{r_{bc}^i\,r_{bc}^j}{r_{bc}^3} )\,p_c^j$ 
involve non-standard classes of integrals that nevertheless have been considered
in the Gaussian basis set with linear terms \cite{lgauss}.

\subsection{Spin variables reduction}

Matrix elements of each spin-independent operator $Q$, after eliminating spin variables, 
takes the standard form
\begin{eqnarray}
\langle \psi'| Q | \psi \rangle &\equiv& \bigl\langle \phi'(r_1,\,r_2,\,r_3)|Q|\, \nonumber\\
&& \times {\cal P} [c_{123}\,\phi(r_1,r_2,r_3)] \bigr\rangle \label{47}
\end{eqnarray}
with $c_{klm}$ coefficients defined in Table~\ref{table2}, and $\cal P$
is the sum of all permutations of $1,2,$ and $3$. This reduction is applicable in an
evaluation of the overlap matrix and the Hamiltonian. Another useful form is obtained for
the Fermi contact matrix element. If we denote
\begin{eqnarray}
\langle \psi'| Q_a|\psi \rangle_F &\equiv& \Big \langle \phi'(r_1,\,r_2,\,r_3)|  \\
&& \times\,{\cal P} \Big[ \sum_a c^{Fa}_{123}\,Q_a\,\phi(r_1,r_2,r_3)\Big] \Big \rangle \nonumber,
\end{eqnarray}
then for the ground state of Be$^+$ with $J=1/2$ and $\vec J = \sum_{a} \vec \sigma_a/2$ we get
\begin{equation}
\frac{1}{J (J+1)}\langle \psi'|\vec J \cdot \sum_a\vec \sigma_a Q_a|\psi \rangle  =  2\, \langle \psi'| Q_a|\psi \rangle_F 
\end{equation}
Second-order terms involve spatially antisymmetric states
\begin{equation}
\psi_A = \frac{1}{\sqrt{6}}\,{\cal A}[\phi(\vec r_1,\vec r_2,\vec r_3)\,\alpha(1)\alpha(2)\,\alpha(3)]\,, 
\end{equation}
for which reduced matrix elements are of the form
\begin{eqnarray}
\langle \psi'_A| Q_{a}|\psi \rangle_A &\equiv& \Big \langle \phi'(r_1,\,r_2,\,r_3)|  \\
&& \times\,{\cal P} \Big[ c^{A}_{123}\,(Q_{1}-Q_{2})\,\phi(r_1,r_2,r_3)\Big]
\Big \rangle \nonumber\\
\langle \psi'_A| Q_{ab}|\psi \rangle_A &\equiv& \Big \langle \phi'(r_1,\,r_2,\,r_3)|  \\
&& \times\,{\cal P} \Big[ c^{A}_{123}\,(Q_{12}-Q_{23})\,\phi(r_1,r_2,r_3)\Big]
\Big \rangle \nonumber
\end{eqnarray}

\subsection{Second-order matrix elements}
Calculations of the second-order terms in Eqs.~\eqref{A6B}, \eqref{A6C} 
and \eqref{A6AN-2} are also highly nontrivial.
The approach using Hylleraas functions encounters 
severe numerical problems. Namely, the optimization of the nonlinear parameters for
the pseudostate in the second-order  matrix elements leads to differences of many 
orders of magnitude between variational parameters, and it destroys the numerical stability 
of the recursion method for extended Hylleraas integrals \cite{recursions2}. 
Also, the complexity of such calculations makes an optimization process very time consuming. 
An alternative solution is the use of a well-optimized ECG basis.
With this, function representations of pseudostates 
can be determined sufficiently accurately and very efficiently. 

At the first step, we reduce spin variables with the help of
a computer algebra program. Next, the second-order elements 
for the ground state of the lithium-like atom involve spatial coordinates only
and are of the following form
\begin{eqnarray}
A^{(6)}_A &=& 4\,\sum_{n\neq 0}\frac{
\langle\psi| [H^A_{a,\rm hfs}]_r|\psi_n\rangle_F\,
\langle\psi_n| [H^A]_r|\psi\rangle}{E-E_n}
\label{redQA}\\
A^{(6)}_B &=& \frac{4}{3}\,\sum_n\frac{
\langle\psi|H^{Bi}_{\rm hfs}|\psi^i_n\rangle\,
\langle\psi^j_n|H^{Bj}_b|\psi\rangle_F}{E-E_n}
\label{redQB}\\
A^{(6)}_C &=&\frac{8}{3}\,\sum_n\frac{
\langle\psi|H^{Cij}_{a,\rm hfs}|\psi^{ij}_{nA}\rangle_A\,
\langle\psi^{kl}_{nA}|H^{Ckl}_{ab}|\psi\rangle_A}{E-E_n}
\label{redQC}
\end{eqnarray}
\begin{table*}[!hbt]
\renewcommand{\arraystretch}{1.1}
\caption{Numerical values of first-order operators in the ground state 
         of Li (Ref.\cite{lit_hfs}) and Be$^+$, H - Hylleraas basis, G - Gausian basis. }
\label{table3}
\begin{ruledtabular}
\begin{tabular}{lcw{4.20}w{4.20}}
Operator  & Basis & \centt{Li} & \cent{{\rm Be}^+}  \\
\hline \\
 $ E = \langle H \rangle                        $ &  H  & -7.478\,060\,323\,910\,10(32)^a & -14.324\,763\,176\,790\,43(22)^a \\
 $ \langle H_A \rangle                          $ &  H  & -12.049\,907\,85(6)           & -43.688\,013\,68(8)   \\
 $ \langle \delta^3(r_a)\rangle_F               $ &  H  &  0.231\,249\,661(2)           &  0.994\,525\,337(5) \\
 $ \langle [\delta^3(r_a)]_{\rm mp}\rangle_F    $ &  H  & -0.027\,726\,521(11)          & -0.087\,880\,92(4)  \\
 $ \langle r_a^{-1} \rangle                     $ &  H  &  5.718\,110\,882\,476\,5(4)   &  7.973\,888\,857\,015\,4(5)  \\
 $ \langle r_a^{-1} \rangle_F                   $ &  H  &  0.360\,344\,320\,41(8)       &  0.628\,135\,118\,56(2) \\
 $ \langle \delta^3(r_a) \sum_{b\neq a} r_b^{-1} \rangle_F $  
                                                  &  H  &  0.419\,203\,4(10)            &  2.620\,526\,3(15)  \\
 $\langle r_a^{-1} \sum_{b\neq a}\delta^3(r_b)  \rangle_F $  
                                                  &  H  &  4.095\,692\,0(4)             &  16.792\,994(4)  \\
 $\langle r_a^{-1}\,\sum_{b>c} \delta^3(r_{bc})\rangle_F  $  
                                                  &  H  &  0.173\,834\,1(2)             &  0.846\,757\,5(3)  \\
 $\langle \delta^3(r_a)\,(\sum_{b \neq a} p_b^2/2 +V + Z\,r_a^{-1}-E )\rangle_F $  
                                                  &  H  &  0.733\,477(4)                &  4.857\,754(5)  \\
 $\langle r_a^{-2} \,(\sum_{b \neq a} \frac{p_b^2}{2}+V +Z\,r_a^{-1}-E) \rangle_F $  
                                                  &  H  &  1.506\,463(3)                &  7.372\,057\,2(8)  \\
 $\langle r_a^{-1}\,( (E-V)^2 - Z^2\,r_a^{-2}) \rangle_F$  
                                                  &  H  &  43.824\,14(2)                &  232.429\,630(7) \\
 $\langle\vec p_a\,r_a^{-2}\,\vec p_a \rangle_F$  
                                                  &  H  &  4.863\,37(4)                 &  28.631\,62(3)  \\[2ex]
 $ E = \langle H \rangle                      $   &  G  & -7.478\,060\,322\,96          & -14.324\,763\,175\,15 \\
 $\langle\sum_{b\neq a}\,\frac{\vec r_{ab}}{r_{ab}^3}\cdot\frac{\vec r_a}{r_a^3} \rangle_F $ 
                                                  &  G  &  0.017\,363\,5(7)             &  0.081\,258\,7(9)  \\
 $\langle\sum_{b\neq a}\,\frac{\vec r_{ab}}{r_{ab}^3}\cdot\frac{\vec r_b}{r_b^3} \rangle_F $  
                                                  &  G  & -0.065\,937\,5(6)             & -0.438\,358\,5(6) \\
 
 $\langle\sum_{b>c} p_b^2\,r_a^{-1}\,p_c^2 \rangle_F $  
                                                  &  G  &  12.663\,6(8)                 &  74.893(4) \\
 
 $\langle\sum_{b>c}p_b^i\,r_a^{-1}\,(
                 \frac{\delta^{ij}}{r_{bc}}+\frac{r_{bc}^i\,r_{bc}^j}{r_{bc}^3} )\,p_c^j \rangle_F $  
                                                  &  G  &  0.266\,794(3)                &  0.922\,84(3)
\end{tabular}
\end{ruledtabular}
\begin{flushleft}
a - Ref.~\cite{slater1} 
\end{flushleft}
\end{table*}
The symmetry of internal pseudostates in the above is determined as follows. 
Since, $[H^A]_r$ is a scalar operator, 
the symmetry of the pseudostate in $A^{(6)}_{A}$ has to be exactly 
the same as that of the wave function in Eq.~\eqref{wf_psiS}. 
For $A^{(6)}_{B}$ and $A^{(6)}_{C}$, the spatial part can be represented  
with elements of $P$-even and $D$ symmetry 
\begin{eqnarray}
\phi_{ab}^{i}(\vec r_1, \vec r_2, \vec r_3) &=& \epsilon_{ijk} r_a^j r_b^k \,\phi(\vec r_1, \vec r_2, \vec r_3), \label{phiPe} \\
\phi_{ab}^{ij}(\vec r_1, \vec r_2, \vec r_3) &=& \bigg(\frac{r_a^i r_b^j}{2} + \frac{r_b^i r_a^j}{2} - 
\frac{\delta^{ij}}{3} \vec r_a \cdot \vec r_b \bigg) \,\phi(\vec r_1, \vec r_2, \vec r_3), \nonumber \\
\label{phiD}
\end{eqnarray}
respectively. The normalization of the corresponding wave functions is set by
Eq. (\ref{47}) with an implicit sum over cartesian indices.
In the calculations of the second-order matrix elements, we generated the ground state wave function 
with ECG basis functions of progressively doubling size from 256 to $2048$ terms. 
Next, for the given external wave function of a given size $N$, the nonlinear parameters 
for pseudostates were optimized using a symmetric second-order element with the corresponding hyperfine
operator. Such a matrix element can be minimized using the adopted variational principle. 
In our approach, the basis set for the pseudostate in $A_A^{(6)}$ is divided into two sectors. 
The first sector is built of the basis functions with the nonlinear 
parameters of size $N/2$ determined in the optimization of the external wave function. 
The nonlinear parameters here are fixed during the optimization 
in order to enable accurate representation of the states orthogonal to the ground state.
The second sector, of size $N$, consists of basis functions that undergo optimization.
For $A_B^{(6)}$ and $A_C^{(6)}$ an orthogonality to the ground state is realized 
by a different symmetry of the basis functions in Eqs.~\eqref{phiPe} and \eqref{phiD}. 
Then, only a single sector is needed with all parameters to be optimized for the basis sizes
$N$ and $2N$, respectively. The size of the pseudostate is chosen to achieve 
convergence for a fixed external wave function.  We noted 
that the symmetric second-order element with $H^{Cij}$ is divergent. Therefore,
in the optimization of the pseudostate for the $A_C^{(6)}$ term
we use a lower singular operator by decreasing the power of $r_a$ by one.  
Due to the more complicated structure of the second-order matrix elements,
both the convergence and the cost of the optimization are less favorable
in comparison to the wave function optimization.

\section{Results}
The final results of the numerical calculations are presented in Table~\ref{table3} and Table \ref{table4}.
The values and their uncertainties have been obtained by extrapolation
of the results obtained for several sizes of basis sets.  
Most of the first-order matrix elements 
in Table~\ref{table3} were obtained in Hylleraas basis sets because of 
the much higher accuracy that can be obtained in comparison to using the ECG functions.
However, some of the matrix elements have been calculated only with
Gaussians. These are the most complicated ones, but they are numerically less
significant than the other terms in $A_N^{(6)}$. The achieved numerical accuracy 
is sufficiently high that the main uncertainty comes from estimation of higher order terms such as those
in Table~\ref{table5}.
All the second-order matrix elements have been calculated only with Gaussian
functions by global optimization of nonlinear parameters in about $1000$
functions, which are used to represent the sum over intermediate states. In spite of the
fact that Gaussian functions do not satisfy the cusp condition at the coalescent
points, they are flexible enough to achieve much greater accuracy
than with Hylleraas functions for second-order matrix elements.

\begin{table}[!hbt]
\renewcommand{\arraystretch}{1.1}
\caption{Numerical values of dimensionless  relativistic and QED corrections to the
  hyperfine splitting in Be$^+$ ion, results from \cite{Yerokhin} in terms of $G_{\rm M1}$ are multiplied by 256/3}
\label{table4}
\begin{ruledtabular}
\begin{tabular}{lw{2.12}}
Contribution  & \centt{Value}  \\
\hline
$A^{(4,0)}$         & 33.326\,863\,92(18) \\
Ref. \cite{YDhfs} & 33.326\,8(8) \\
$A^{(4,1)}$         & 97.035\,673\,8(13) \\
Ref. \cite{YDhfs} & 102.(18.) \\[1ex]
 $A^{(6)}_{A}$ &   1\,196.97(4) \\
 $A^{(6)}_{N}$ &     -133.631(16) \\
 $A^{(6)}_{B}$  &        0.134\,6(2) \\
 $A^{(6)}_{C}$  &        0.814\,2(6) \\
 $A^{(6)}_{R}$  &     -240.866\,95 \\[1ex]
 $A^{(6)}$      &      823.42(4)  \\
 Ref. \cite{Yerokhin} & 823. \\
 Ref. \cite{YDhfs}    & 756.(42.) \\
 $A^{(7)}$      &    -2\,992.(481)
\end{tabular}
\end{ruledtabular}
\end{table}
Results of the expansion of (dimensionless) hyperfine constant ${\cal A}$ in powers of $\alpha$
and the mass ratio for the Be$^+$ ion are presented in Table \ref{table4}. Leading order
terms, the Fermi contact interaction, and the mass polarization correction
are compared with previous results by Z. -C. Yan {\em et al.} \cite{YDhfs}. 
Our result for $A^{(6)}$ are in excellent agreement with the relativistic CI
calculations of Yerokhin \cite{Yerokhin}. It gives us confidence
in the theoretical approach and in the numerical results obtained in this work. 
\begin{table}[!hbt]
\renewcommand{\arraystretch}{1.1}
\caption{Contributions in MHz to the hyperfine splitting constant $A$ in
  $^{9}$Be$^+$, physical constants are  $g=2.002\,319\,304\,361\,53(53)$,
  $\alpha^{-1} = 137.035\,999\,074(44) $. 
 The second uncertainty of $A_{\rm the}$ comes from the nuclear magnetic moment.}
\label{table5}
\begin{ruledtabular}
\begin{tabular}{lw{4.15}} & \centt{$^9$Be$^+$} \\
\hline
 $\mu[\mu_N]$  Ref. \cite{itanoBe9}          & -1.177\,432(3)        \\
 atomic mass $[u]$ Ref. \cite{nu_mass}       & 9.012\,182\,20(43)     \\
 $g_{\rm N}$                                  &  -1.755\,335\,5(25)    \\
 $\varepsilon \times 10^{-9}$                &  -6.602\,679(17)        \\
 $\varepsilon\,\alpha^4\,g/2\,A^{(4)}$       &  -624.600\,44           \\
 $\varepsilon\,\alpha^5\,A^{(5)}_{\rm rec}$    &     0.006\,85           \\
 $\varepsilon\,\alpha^6\,A^{(6)}$            &    -0.820\,96(4)         \\
 $\varepsilon\,\alpha^7\,A^{(7)}$            &     0.021\,8(36)        \\[1ex]
 $A_{\rm the}$ (point nucleus)                &  -625.392\,7(36)(16)    \\
 Ref. \cite{Yerokhin}                       &  -625.401(22)           \\
 $A_{\rm exp}$ Ref. \cite{winelandBe9}        &  -625.008\,837\,048(10) \\[1ex]
  $(A_{\rm exp}-A_{\rm the})/A_{\rm exp}$         & \multicolumn{1}{c}{$-614(6)(3)   $ ppm\phantom{0}} \\
 Ref. \cite{Yerokhin} (theory)       & \multicolumn{1}{c}{$-514(16)$ ppm} \\
 $\tilde r_Z$                               & 4.07(5)(2) \mathrm{\;fm} \\
 $r_{E}$ Ref. \cite{jansen}                      & 2.519(12)  \mathrm{\;fm}  \\
  \end{tabular}
\end{ruledtabular}
\end{table}
Table \ref{table5} summarizes the results for the $^9$Be$^+$ ion.
From the measured hyperfine constant and the magnetic moment,
we determined finite nuclear size effects, which are expressed
in terms of the effective Zemach radius $\tilde{r}_Z$. We observed
that the experimentally determined $\tilde r_z(^{9}{\rm Be})$ does not
agree well with the approximate nuclear structure calculations in Ref. \cite{Yerokhin}.
Bearing in mind the significant differences in $\tilde r_z$ in Li isotopes,
considerable theoretical work is needed to correctly describe
the finite nuclear size and polarizability effects in the atomic hyperfine splitting. 

\section{Summary}
We have developed a nonrelativistic QED approach to
the hyperfine splitting in light atomic systems
and have demonstrated that from the comparison to experimental values
one can obtain valuable information about the finite nuclear distribution.
We observed in Ref. \cite{lit_hfs} that the Zemach radius for $^6$Li is about 40\%
smaller than that of $^7$Li. Here we demonstrate 
that by means of atomic spectroscopy one can obtain the Zemach
radius for Be isotopes, and give example for $^9$Be, for which the magnetic moment
is  well known. For other Be isotopes, although hyperfine splitting is
known \cite{okadaBe7, takamineBe11}, the magnetic moment has not yet been measured
with the sufficient accuracy. 
We do not attempt here to accurately relate $\tilde{r}_Z$ to the distribution of the
magnetic moment, as our knowledge of nuclear theory is not sufficient.
We point out, however, that this model is an independent and very accurate
(as accurate as the magnetic moment) method to approach nuclear
magnetic moment distribution.

\section*{Acknowledgments}
The authors acknowledge support from NCN Grants 2012/04/A/ST2/00105 and 2011/01/B/ST4/00733.



\begin{thebibliography}{99}
\bibitem{litiso} W. N{\"o}rtersh{\"a}user, {\it et al.}, Phys. Rev. A 83, 012516 (2011).
\bibitem{lit_hfs} M. Puchalski, and K. Pachucki, Phys. Rev. Lett. 111, 243001 (2013).
\bibitem{krp_hfs} K. Pachucki,  Phys. Rev. A {\bf 66}, 062501 (2002). 
\bibitem{nist} Peter J. Mohr, Barry N. Taylor, and David B. Newell, Rev. Mod. Phys. {\bf 84}, 1527 (2012).
\bibitem{eides} M.I. Eides, H. Grotch, and V.A. Shelyuto, Phys. Rep. {\bf 342}, 63 (2001).
\bibitem{UJ} B. J. Wundt and U. D. Jentschura, Phys. Rev. A {\bf 83}, 052501 (2011).
\bibitem{lit_wave} M. Puchalski and K. Pachucki, Phys. Rev. A {\bf 73}, 022503 (2006).
\bibitem{king_lit} F. W. King, J. Mol. Struct.: THEOCHEM 400, 7 (1997).
\bibitem{yan_lit}  Z.-C. Yan and G.W.F. Drake, Phys. Rev. A {\bf 52}, 3711 (1995);
         Z.-C. Yan, M. Tambasco, and G.W.F. Drake, Phys. Rev. A {\bf 57}, 1652 (1998);
         Z.-C. Yan, W. N\"ortersh\"auser and G.W.F. Drake, Phys. Rev. Lett. {\bf 100}, 243002 (2008); 
         ibid. {\bf 102}, 249903(E) (2009).
\bibitem{Wang_2012} L. M. Wang, Z.-C. Yan, H. X. Qiao, and G. W. F. Drake, Phys. Rev. A {\bf 85}, 052513 (2012).
\bibitem{slater1} M. Puchalski, D. Kedziera, and K. Pachucki Phys. Rev. A {\bf 80}, 032521 (2009).
\bibitem{recursions} K. Pachucki, M. Puchalski and E. Remiddi, Phys. Rev. A {\bf 70}, 032502 (2004).
\bibitem{recursions2} K. Pachucki and M. Puchalski, Phys. Rev. A {\bf 71}, 032514 (2005);
                     Phys. Rev. A {\bf 77}, 032511 (2008).
\bibitem{lit_rel} M. Puchalski and K. Pachucki, Phys. Rev. A {\bf 78}, 052511 (2008).
\bibitem{dlines} M. Puchalski, D. Kedziera,  and K. Pachucki, Phys. Rev. A {\bf 87}, 032503 (2013) 
\bibitem{king_int} P. J. Pelzl, G. J. Smethells, and F. W. King, 
                   Phys. Rev. E {\bf 65}, 036707 (2002);
                   D. M. Feldmann, P. J. Pelzl and F. W. King,
                   J. Math. Phys. {\bf 39}, 6262 (1998).
\bibitem{gamma1} R.A. Sack, C.C.J Roothaan and W. Ko\l os, J. Math. Phys. {\bf 8}, 1093 (1967).
\bibitem{gamma2} V.I. Korobov, J. Phys. B {\bf 35}, 1959 (2002).
\bibitem{gamma3} F.E Harris, A.M. Frolov and V.H. Smith, Jr., J. Chem. Phys {\bf 121}, 6323 (2004).
\bibitem{mitroy2013} J. Mitroy, S. Bubin, W. Horiuchi, Y. Suzuki, L. Adamowicz, W. Cencek, K. Szalewicz, J. Komasa, D. Blume, and K. Varga, 
Rev. Mod. Phys. {\bf 85}, 693 (2013).
\bibitem{slater2} M. Puchalski and K. Pachucki, Phys. Rev. A 81, 052505 (2010).
\bibitem{lgauss} K. Pachucki and J. Komasa, Chem. Phys. Lett. {\bf 389}, 209 (2004); 
                 Phys. Rev. A  A {\bf 70}, 022513 (2004).
\bibitem{Yerokhin} V. A. Yerokhin, Phys. Rev. A {\bf 78}, 012513 (2008).
\bibitem{YDhfs} Z.-C. Yan, D. K. McKenzie, and G. W. F. Drake, Phys. Rev. A {\bf 54}, 1322 (1996).
\bibitem{itanoBe9}  W. M. Itano, Phys. Rev. B 27, 1906 (1983).
\bibitem{nu_mass} G. Audi, A.H. Wapstra, and C. Thibault, Nucl. Phys {\bf A729}, 337 (2003).
\bibitem{winelandBe9} D. J. Wineland, J. J. Bollinger, and W. M. Itano, Phys. Rev. Lett. {\bf 50}, 628 (1983).
\bibitem{jansen} J.A. Jansen, R.Th. Peerderman, and C. deVries, Nucl. Phys. A {\bf 188},337 (1972).
\bibitem{okadaBe7} K. Okada, {\em et al.} Phys. Rev. Lett. {\bf 101}, 212502 (2008).
\bibitem{takamineBe11} A. Takamine, {\em et al.}, submitted to Phys. Rev. Lett.
\end{thebibliography}
\end{document}